# Two low-power optical data transmission ASICs for the ATLAS liquid argon calorimeter readout upgrade

**Le Xiao**[a,b]**, Wei Zhou**[a,b]**, Quan Sun**[b]**, Binwei Deng**[c]**, Datao Gong**[b,1]**, Di Guo**[b]**, Huiqin He**[d]**, Suen Hou**[e]**, Chonghan Liu**[b]**, Tiankuan Liu**[b]**, James Thomas**[b]**, Jian Wang**[b,f]**, Annie C. Xiang**[b]**, Dongxu Yang**[b,f]**, Jingbo Ye**[b]**, Xiandong Zhao**[b]

[a] *Department of Physics, Central China Normal University,*
*Wuhan, Hubei 430079, P.R. China*

[b] *Department of Physics, Southern Methodist University,*
*Dallas, TX 75275, USA*

[c] *Hubei Polytechnic University,*
*Huangshi, Hubei 435003, P.R. China*

[d] *Shenzhen Polytechnic,*
*Shenzhen 518055, P.R. China*

[e] *Institute of Physics, Academia Sinica,*
*Nangang 11529, Taipei, Taiwan*

[f] *State Key Laboratory of Particle Detection and Electronics, University of Science and Technology of China,*
*Hefei Anhui 230026, P.R. China*

E-mail: dgong@mail.smu.edu

ABSTRACT: A serializer ASIC and a VCSEL driver ASIC are needed for the front-end optical data transmission in the ATLAS liquid argon calorimeter readout phase-I upgrade. The baseline ASICs are the serializer LOCx2 and the VCSEL driver LOCld, designed in a 0.25-µm Silicon-on-Sapphire (SoS) CMOS technology and consumed 843 mW and 320 mW, respectively. Based on a 130-nm CMOS technology, we design two pin-to-pin-compatible backup ASICs, LOCx2-130 and LOCld-130. Their power consumptions are much lower then of their counterparts, whereas other performance, such as the latency, data rate, and radiation tolerance, meet the phase-I upgrade requirements. We present the design of LOCx2-130 and LOCld-130. The test results of LOCx2-130 are also presented.

---

[1] Corresponding author.

# Contents

## 1. Introduction

Optical data links have been successfully providing reliable and efficient data transmission in Large Hadron Collider (LHC) experiments for the past decade [1-2]. When LHC upgrades to high luminosities, new readout electronics for ATLAS are required for the trigger system to improve the events selection algorithm[3]. In the ATLAS Liquid Argon (LAr) trigger electronics upgrade, the L1-trigger is removed from the front-end board. The data rate of the trigger board increases by a factor of 400 to about 200 Gbps per board. The latency of the data transmitted from the detector to the control room is required to be not greater than 150 ns (not including the time passing through the optical fiber). The power consumption budgets of the optical link is also limited to no greater than 100 mW per Gbps [3]. The electronic system on the front-end board must be radiation tolerant [4-5]. In order to meet these requirements, a high-speed, low-power and low-latency optical link is needed for the ATLAS LAr Calorimeter trigger system upgrade.

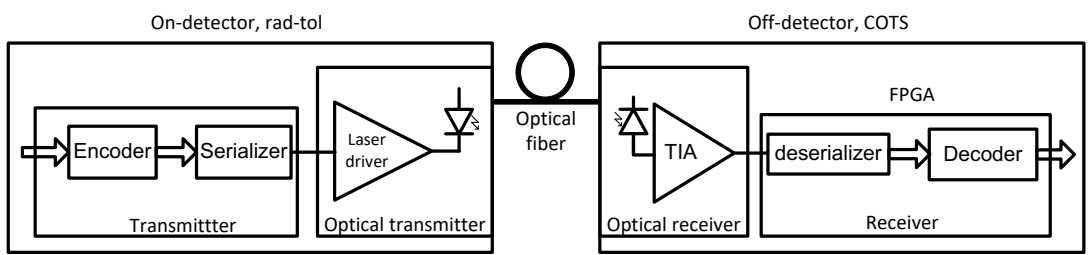

**Figure 1:** Block diagram of optical data transmission system.

Figure 1 is a generic block diagram of the optical link transmitting data from front-end to back-end in the high-energy experiment. The transmitter, integrates an encoder and a serializer, encodes the data coming from the upstream Analog-to-Digital Converters (ADCs), and provides a serial output. The encoder scrambles the data and builds a data frame. The optical transmitter converts the electrical signal to an optical signal. The optical signal is recovered to an electrical signal by an optical receiver in the counting room. The deserializer, usually embedded in a Field-Programmable Gate Array (FPGA), recovers the parallel data from the serial data. An Application Specific Integrated Circuit (ASIC) transmitter is needed for the front-end optical readout due to the demand on data bandwidth, high channel density, low power consumption, low transmission latency, and radiation tolerance [4-5]. For an optical link system, Total Ionizing Dose (TID), Single Event Upset (SEU), and Single Event Latchup (SEL) effects are considered to ensure the correct data transmission. The new LAr Trigger Digitizer Boards (LTDB) [6-7] also requires that the Bunch Crossing Identification (BCID) is encoded in the data frame for data alignment at the backend. The latency requirement for the transmitter, including the encoder and serializer, is 75 ns, which is beyond the GigaBit Transceiver ASIC (GBTX) [8] specification.

Because the 0.25 um Silicon-on-Sapphire process is phasing out of the market, we need to re-design the serializer and the laser driver in a mature commercial process. Based on a commercial 130-nm CMOS process of Global Foundry, we have designed a transmitter ASIC, LOCx2-130, and a Vertical Cavity Surface Emitting Laser (VCSEL) driver ASIC, LOCld-130, to meet these requirements as a pin-to-pin compatible backup of LOCx2 [9] and LOCld [10], for the ATLAS Liquid Argon Calorimeter Phase-I trigger upgrade. The meaning of the abbreviation LOC used in those ASICs is the Link on Chip. LOCx2-130 is re-designed based on the GBTX analog core. LOCld-130 is a new design. GBTX is a general Serializer/Deserializer (SERDES) chip which has a general interface with large latency, greater than the requirements in a triggering system. LOCx2-130 has designed a low-latency and low-power digital interface. LOCx2-130 is a two-channel transmitter ASIC. Each channel receives data from the upstream ADCs and encodes the data, then the encoded data will be serialized and transmitted at 4.8 Gbps. LOCld-130 is a low-power two-channel VCSEL driver ASIC and each channel operates at 5.12 Gbps.

The remainder of the paper is organized as follows: Section 2 describes the design of LOCld-130. Section 3 discusses the design of LOCx2-130. The test setup and measurement results of LOCx2-130 are presented in Section 4. Section 5 summarizes the paper.

## 2. Design of LOCld-130

LOCld-130 is a drop-in backup of LOCld. Compared with LOCld, LOCld-130 has a simplified interface, simpler channel structure, and lower power consumption. The power supply of LOCld-130 is 1.5 V, the minimum input signal to LOCld-130 is 200 mV, and the data rate is up to 5.12 Gbps. The LOCld-130 is composed of an analog core and an Inter-Integrated Circuit (I2C) slave module. The block diagram of LOCld-130 is shown in Figure2. The analog core has a two-stage limiting amplifier (LA) and a high-current differential driver. Each channel in the ASIC is individually powered, making the ASIC suitable for applications in dual-channel transmitters and in transceivers.

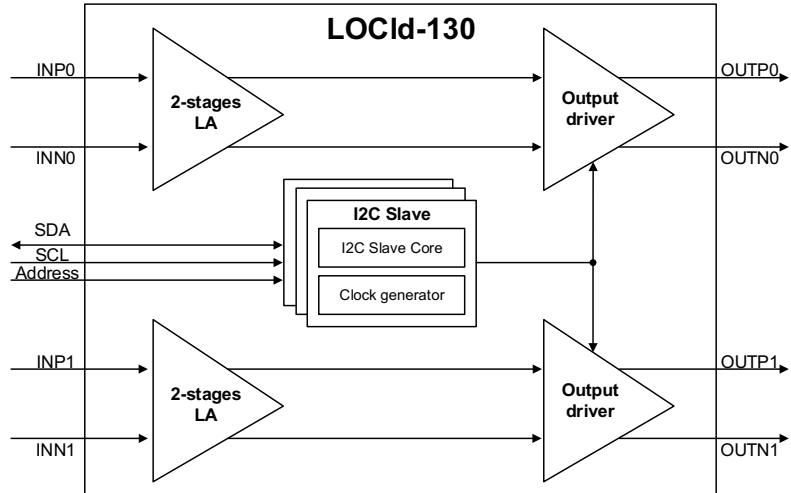

**Figure 2:** The block diagram of LOCld-130

The schematic of LA is shown in Figure 3, a two-stage amplifier is chosen to optimize the gain as well as the bandwidth. The bias current of each stage is 1.5 mA and 3 mA respectively. A 3.2-nH peaking inductor is shared between the two stages, increasing the bandwidth of the LA to more than 3.5 GHz, which causes negligible inter-symbol interference (ISI) jitter for a 5-Gbps signal [11]. The LA gain is 14 dB, amplifying input signal from 200 mVp-p to 1 Vp-p. The output signal of the LA is designed to fully steer the current of the last differential stage from one arm to the other one to maximize the electro-optical efficiency. The output signal voltage swing of LA is set at 1 Vp-p, two thirds of the power supply voltage.

The output driver adopts a high-current differential structure which pull-up 50 ohm resistance for impedance matching. The modulation current provided by the output driver is adjustable through a 6-bit current Digital to analog converter (DAC). The range of modulation current for VCSEL is from 2 mA to 10 mA. An optical driver on the front-end is required to work properly at power up so that the backend can configure the front-end board conveniently. At the default configuration, the driver outputs an 8-mA modulation current and a 3-mA bias current, which ensures the VCSEL diode works after power up. The driver may not be in an optimal state, but the control room can re-configure both the bias current and the modulation current after power up.

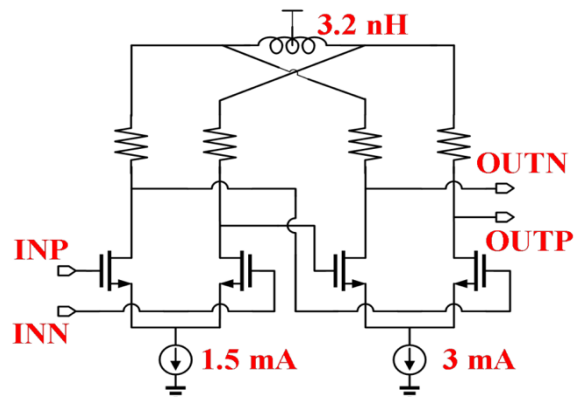

**Figure 3:** The schematic of the LA for LOCld-130

The I2C slave is composed of a clock generator and an I2C slave core. The clock generator is a low-frequency ring oscillator which is made up of many inverter cells and delay cells from the digital standard cell library. The clock generator produces the clock for the I2C slave core only during communication, which means the whole I2C has no dynamic power consumption after the configuration is completed. The whole layout and the pin map of the clock generator and the I2C slave core have been generated through the digital design flow. In LOCld-130, only the I2C slave has memory cells. In order to be immune to SEU, both the clock generator and the I2C slave core are protected with Triple Modular Redundancy (TMR).

The die area of LOCld-130 is 1.0 mm × 2.8 mm. Figure 4 is the layout of LOCld-130. The analog cores of two channels, including the LA and the output high-current differential driver, are located at the left and the right of the floor plan respectively. The digital functional blocks, including the clock generator and the I2C slave core located at the middle of the floor plan, are noisier than the analog core. We carefully isolated the substrates of these circuits and provide separated power supply and ground for each of them.

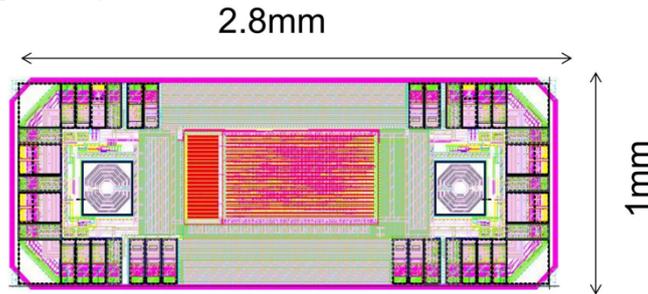

**Figure 4:** Layout of LOCld-130

The whole channel post-layout simulation with Pseudo Random Binary Sequence 7 (PRBS7) input is conducted. The load model of VCSEL, bonding wire inductor introduced in the package, and the noise from the power supply have been taken into account in the simulation. The eye diagram of the current to the laser in typical configuration is shown in Figure 5. The latency of LOCld-130 is less than 200 ps. The deterministic jitter in the simulation is less than 5 ps. If the random jitter caused by the power supply noise and transistor noise is considered, the total jitter is not more than 30 ps. The power consumption of LOCld-130 is about 56 mW per channel. We compare LOCld-130 with GigaBit Laser Driver (GBLD) [12] and LOCld in table 1. Because GBLD drives both VCSEL and edge-emitting laser (EEL) diode, it consumes more power.

Table 1. A comparison of LOCld-130 with LOCld and GBLD

|  | GBLD | LOCld | LOCld-130 |
|---|---|---|---|
| Functions | VCSEL/EEL driver | VCSEL driver | VCSEL driver |
| Number of channels | 1 | 2 | 2 |
| Process | 130 nm Bulk CMOS | 0.25 µm Silicon-on-Sapphire CMOS | 130 nm Bulk CMOS |
| Data rate (Gbps) | 4.8 | Up to 8.0 | 5.12 |
| Power consumption (mW) | 500 | 160 × 2 | 56 × 2 |
| Total Jitter(ps) | <40 | <30 | <30 |

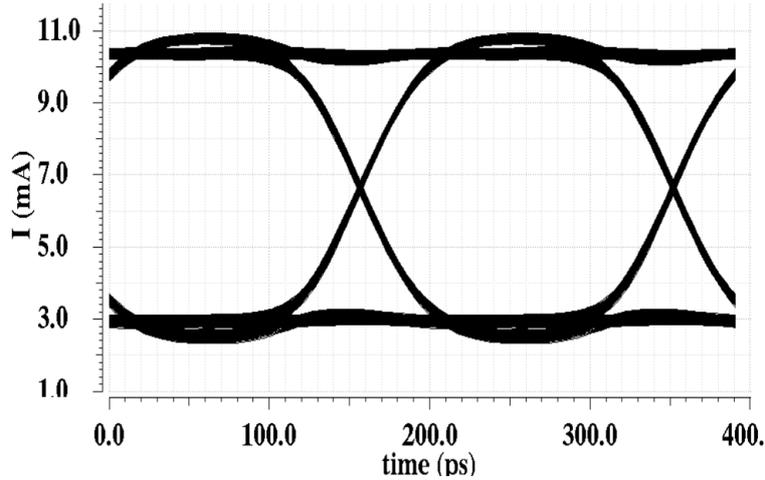
**Figure 5:** Typical eye diagram of LOCld-130 at 5.12 Gpbs

## 3. Design of LOCx2-130

LOCx2-130 supports three types of ADCs, an ASIC called Nevis ADCs [13] and two Commercial-Off-The-Shelf (COTS) devices (part numbers ADS5272 [14] and ADS5294 [15] produced by Texas Instruments). The input data of each channel of LOCx2-130 come from two Nevis ADCs, one ADS5272, or one ADS5294. LOCx2-130 implements a custom line code called LOCic [16]. The output frame definition of the LOCx2-130 is shown in Figure 6.

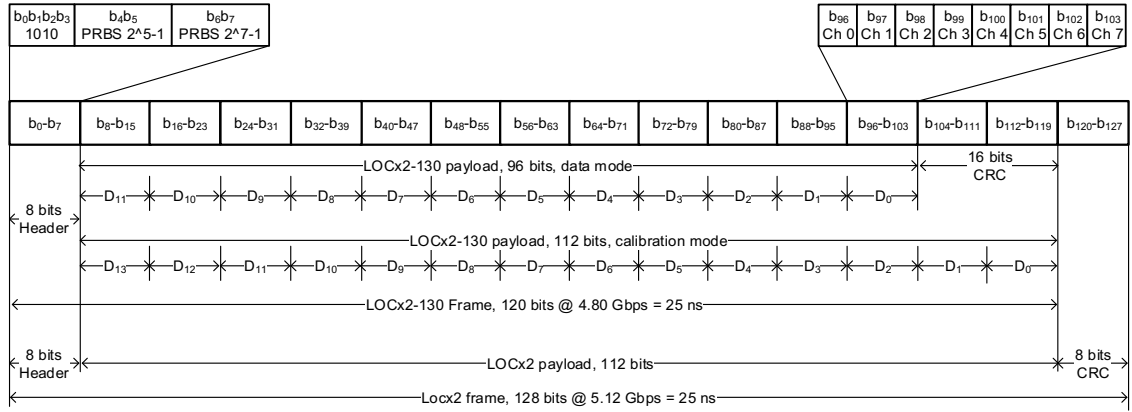
**Figure 6:** LOCic frame definition.

The data frame also transfers BCID information (from 0 to 3563) with low overhead. The 120-bit length data frame includes an 8-bit frame header, 96-bit payload and 16-bit frame trailer. The frame header identifies the frame boundary and also efficiently transfers the BCID for each frame. In a normal data transition mode, the 16-bit frame trailer is a 16-bit Cyclic Redundant Checking (CRC) code calculated from the payload to detect the potential transmission errors. It also can be set in the calibration mode in which the CRC code is replaced by 16-bit more payload. Therefore, a data frame carries 14-bit ADC data. The ADS5294 can only work in this mode since it is a 14-bit ADC. The payload is scrambled to keep the signal DC-balanced in serial data transmission. Neither the frame header nor the frame trailer is scrambled.

We utilize the GBTX analog core, which has been modified to a 30:1 serializer at 4.8 Gbps and named as TDS [17]. The frequency of the digital circuit frequency of TDS is 160 MHz, 4 times of the parallel digital circuits of the original GBTX. The high frequency speeds up the data interface and the encoding process and reduces the latency. Although the data frame size is 120 bits, the encoding process is optimized in a pipeline structure to minimize the latency. The block diagram of LOCx2-130 is shown in Figure 7. The LOCx2-130 is composed of two encoders, two 30:1 serializers, two output drivers, a shared Phase-Locked-Loop (PLL), and an I2C slave. The encoder, named as LOCic-130, encodes the ADC data for the serializer. The PLL provides multiple clock signals up to 2.4-GHz for each serializer. The $I^2C$ slave is used for the GBTX-based control link [18] to configure the chip. LOCx2-130 requires a 40-MHz LHC reference clock and a BCID reset, which both come from the GBTX-based control link. The serialized data are sent to an optical module named MTx [19]. LOCx2-130 reuses the silicon-proven components as much as possible. We modify the analog core so that two serializers share a single PLL to reduce the power consumption. The latency of the LOCx2-130 is power-cycle independent.

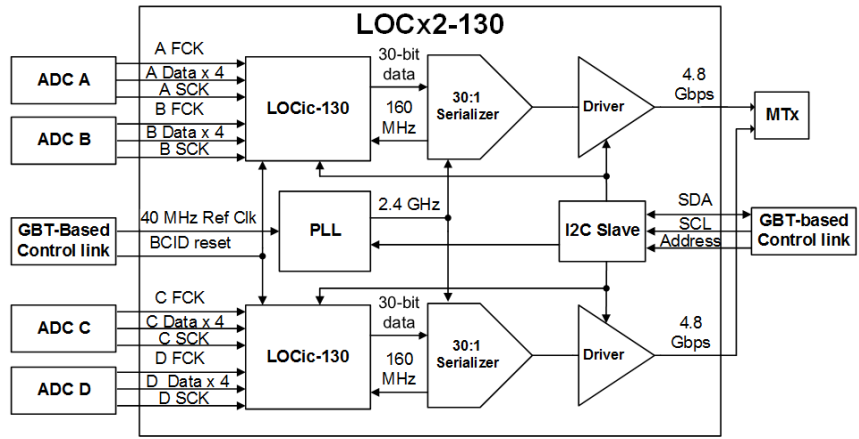

**Figure 7:** The block diagram of LOCx2-130.

The LOCic-130 encoder diagram is shown in Figure 8. The LOCic-130 encoder is composed of an ADC interface, a synchronous First-In-First-Out (FIFO), and a core encoder. All three types of ADCs have the same output signals (Data, Frame, and data Clock) but have different channel number, level, timing, and format specifications. The ADC interface receives data from an ADS5272, a ADS5294, or two Nevis ADCs and provides unified outputs for the FIFO. The FIFO accommodates the different data rates of three types of ADCs. The core encoder builds the data frame and outputs to the serializer.

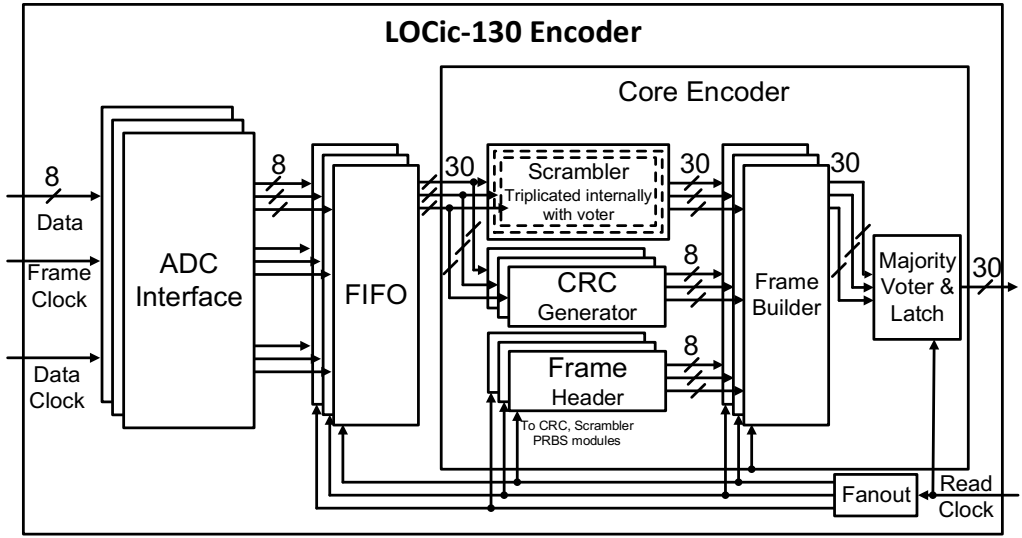

**Figure 8:** The block diagram of LOCic-130.

The encoder builds the data frame and outputs 30 bits of data to the serializer at 160 MHz. The encoder includes a CRC generator, a scrambler, a frame header generator, and a frame builder. The CRC code is calculated from the unscrambled payload using $x^{16}+x^{14}+x^{12}+x^{11}+x^{9}+x^{8}+x^{7}+x^{4}+x+1$. The CRC code detects all odd bit flips or even bits up to 4. For other even bit flips, the probability that the CRC code misses the error is about 0.003%. The scrambler is used to generate DC-balanced codes without introducing extra bits. The scrambling polynomial, $x^{58} + x^{39} + 1$, is widely used in industry [20]. The LOCic-130 frame header generator produces a fixed 1010 pattern followed by the 4-bit BCID field. The frame builder assembles the data into 30-bit width data and feeds the data to the 30:1 serializer.

The LOCic-130 is protected by triple redundant method to be immune to SEU. The LOCic-130 TMR structure diagram is shown in Figure 8. When implementing the TMR, the modules that have a pipeline structure without internal feedback or have internal feedback but are reset periodically, are simply instantiated three times to triple redundancy. The Scrambler, however, has feedback and no reset mechanism. It is triplicated internally with the voter added at the input of every D flip-flop to eliminate error propagation. A final Majority Voter and Latch is added after the triplicated Frame Builder. So, the outputs of LOCic-130 are 30-bit voted parallel data. Since it is a quick design we have not designed a SEU monitor in this chip to get the SEU rate which is useful to understand the radiation environment. To prevent the SEL, guard rings are carefully laid out in the design.

The die area of LOCx2-130 is 2.0 mm × 5.0 mm. Figure 9 is the layout of LOCx2-130. LOCx2-130 is packaged in a 100-pin plastic quad-flat no-leads (QFN) package. A picture of two packaged chips is shown in Figure 10.

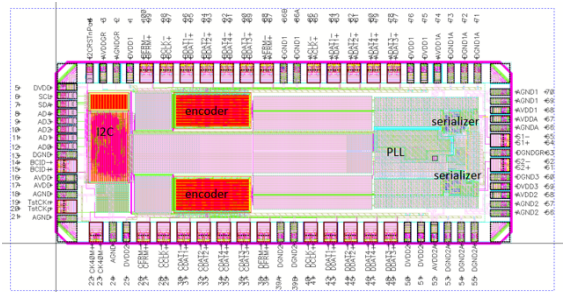 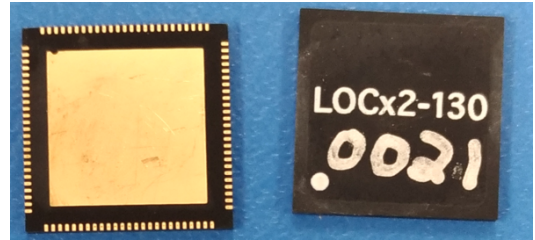

**Figure 9**. Layout of LOCx2-130.     **Figure 10.** QFN packaged LOCx2-130 ASIC**.**

## 4. Test results of LOCx2-130

LOCx2-130 has been evaluated in lab tests. The block diagram of the test setup is shown in Figure 11. A clock board (Si5338 Evaluation Board Kit produced by Silicon Labs) generates three synchronized differential clock signals for the test system. A 50-GSample/s real-time oscilloscope (Model DSA 72004 produced by Tektronix) is used to observe eye diagrams and measure the jitter. A Xilinx Kintex-7 FPGA KC705 Evaluation Kit (Part number EK-K7-KC705-G produced by Xilinx) is used to measure the Bit Error Rate (BER) and link latency. The FPGA emulates ADCs as the input of the LOCx2-130 test board. The FPGA generates a BCID reset signal, which lasts for one cycle and repeats every 3564 cycles of the 40-MHz clock. The FPGA also implements two link receivers. The link receiver includes the deserializer of a Multiple-Gigabit Transceiver (MGT), a LOCic-130 decoder, and an error logger. The error logger status is monitored on a personal computer via the Xilinx ChipScope Pro tool. The serial-data outputs of LOCx2-130 are sent to either the high-speed real-time oscilloscope or the FPGA. We use an I$^2$C master (Model USB-8451, Part No. 779553-01, produced by National Instruments) to configure the LOCx2-130 in the test.

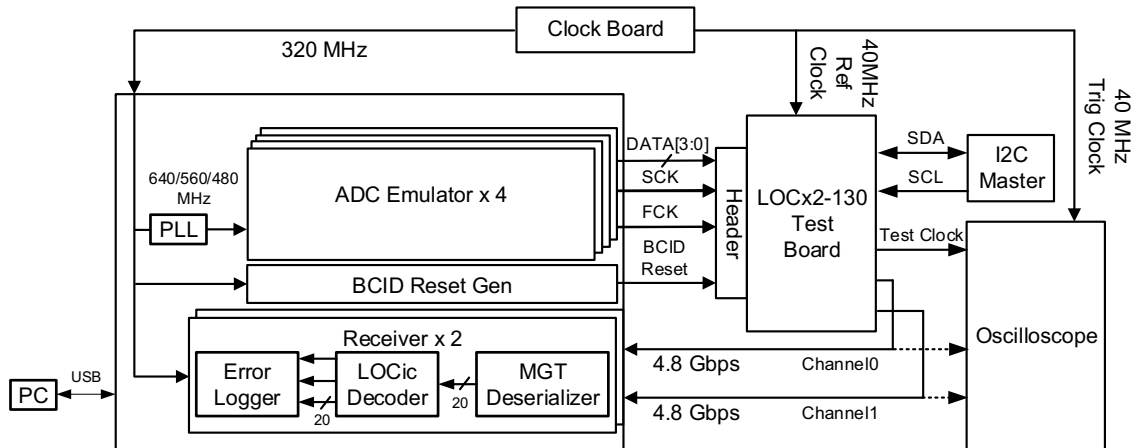

**Figure 11:** The block diagram of test setup.

The eye diagram of LOCx2-130 at 4.8 Gbps is shown in Figure 12. The rise time and the fall time are about 78 ps. At the data rate of 4.8 Gbps, the deterministic jitter is 28 ps (peak-peak) and random jitter is 2.3 ps (RMS), corresponding to total jitter of less than 52 ps (peak-peak) at the bit error rate of $10^{-12}$. The typical differential amplitude of the output is 300 mVp-p.

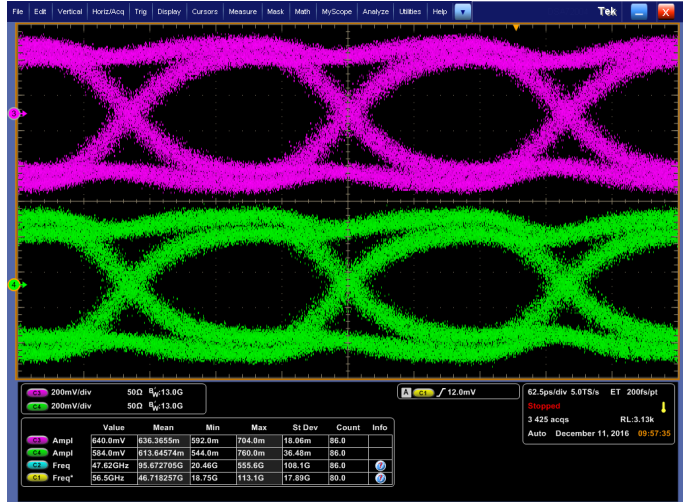
**Figure 12:** An eye diagram of LOCx2-130 at 4.8 Gbps.

The LOCx2-130 transmitter is fully functioning in the test. The BCID information was completely recovered in the decoder and the CRC code regenerated from the received data was consistent with the received CRC code. Moreover, the received payloads were consistent with that generated in the ADC emulator. We tested a link system with LOCx2-130 for a week to check the stability and no single error occurred in the data transmission. We also tested 20 chips for 16 minutes each. No error was observed during the test for 16 minutes, meaning that the bit error rate was less than $10^{-12}$ at the confidence level of 99% [21] for all these chips.

The latency of the whole link, including LOCx2-130 and the receiver implemented in an FPGA, not including optical transceiver and the optical fiber, was measured. The latency of the whole link is no more than 101.6 ns, achieving the design goal of no more than 150 ns. The latency contribution of LOCx2-130 is 34.4~40.7 ns. The latency variation is much less than an LHC clock cycle. When the data are latched with a 40-MHz LHC clock on the receiver side and then sent to the trigger system, the latency is still fixed.

Table 2 compares the performance of GBTX, LOCx2 and LOCx2-130. As can be seen in the table, both LOCx2 and LOCx2-130 meet the requirements of LAr phase-I upgrade, but LOCx2-130 consumes only 440 mW, much less than LOCx2.

Table 2. A comparison of LOCx2-130 with LOCx2 and GBTX

|  | GBTX | LOCx2 | LOCx2-130 |
| --- | --- | --- | --- |
| Function | transceiver | dual-channel transmitters | dual-channel transmitters |
| Process | 130 nm Bulk CMOS | 0.25 μm Silicon-on-Sapphire CMOS | 130 nm Bulk CMOS |
| BCID embedded | No | Yes | Yes |
| Latency (ns) | 212.5 | 27.3 | 40.7 |
| Data rate (Gbps) | 4.8 | 5.12 | 4.8 |
| Power consumption(W) | 2.2 | 0.84 | 0.44 |

We also tested LOCx2-130 using x-rays. The total dose reached 3.0 kGy, exceeding the requirements of the ATLAS Lar Calorimeter phase-I upgrade, and we did not observe significant changes in the power consumption or eye diagrams. We also tested this chip in a 200-MeV proton beam. No single SEU or SEL event was observed when the fluence was accumulated to $9\times10^9$ p/cm$^2$. We calculated that the bit error rate in the application is less than $3.9\times10^{-15}$ at the 99% confidence level.

## 5. Conclusion

We present the design and test results of LOCx2-130, a serializer ASIC for the ATLAS Liquid Argon Calorimeter trigger upgrade. LOCx2-130 consists of two serializer channels and each channel encodes ADC data and transmits serial data at 4.8 Gbps with a latency of less than 40.7 ns. The power consumption of LOCx2-130 is about 440 mW. We also present the design of LOCld-130, a 5.12 Gbps VCSEL driver ASIC designed as a drop-in backup to LOCld. The total power consumption of LOCld-130 is 112 mW (56 mW/channel) including the VCSELs. Both chips meet the requirements of this application and the power consumptions are much lower then the previous version.

## Acknowledgments

We acknowledge the support by the NSF and the DOE Office of Science, SMU's Dedman Dean's Research Council Grant, and the National Natural Science Foundation of China under Grant No. 11705065. We are grateful to Dr. Paulo Moreira, Syzmon Kulis, and Sandro Bonacini from CERN; Drs. Jinhong Wang and Junjie Zhu from the University of Michigan; Drs. Hucheng Chen, Kai Chen, and Hao Xu of Brookhaven National Laboratory; Dr. Nicolas Dumont Dayot of LAPP, and Dr. Bernard Dinkespiler of CPPM.